# Security and Privacy of Wireless Beacon Systems

*Aldar C-F. Chan[1], Raymond M. H. Chung[2]*

***Abstract.*** *Bluetooth Low Energy (BLE) beacons have been increasingly used in smart city applications, such as location-based and proximity-based services, to enable Internet of Things to interact with people in vicinity or enhance context-awareness. Their widespread deployment in human-centric applications makes them an attractive target to adversaries for social or economic reasons. In fact, beacons are reportedly exposed to various security issues and privacy concerns. A characterization of attacks against beacon systems is given to help understand adversary motives, required adversarial capabilities, potential impact and possible defence mechanisms for different threats, with a view to facilitating security evaluation and protection formulation for beacon systems.*

## 1. Introduction

BLE beacons, or simply, beacons, are small, low-cost radio transmitting devices which are short-range and usually mounted at strategic locations or attached to moving objects for identifying locations or tracking objects respectively. BLE beacons, with low power consumption and high-quality proximity estimation features, have demonstrated great potential in smart city applications, especially, in context-aware, location-based and proximity-based services. Smart and social objects [2] and contact tracing are some application examples. A beacon, based on the advertising mechanism of the BLE wireless communication protocol [3], regularly broadcasts short messages of static data (which is usually its beacon identity or ID) to advertise its presence to BLE-enabled devices (such as smartphones) in proximity. Pre-installed applications on these nearby devices — upon picking up the beacon messages — can then trigger actions like pulling context-aware information or notifications to the devices. The specific actions triggered are usually relevant to the physical location or context [8, 14]. Unlike other BLE devices, typical beacons only support one-way communication.

Since beacons are typically deployed in human-centric IoT (Internet of Things) and smart city applications involving users' everyday activities that depend on fine-grained location information, ensuring security and privacy of beacon systems is crucially important. Yet, beacon systems face various kinds of attacks, with security and privacy issues [4-6, 9-12, 13-14]. The widespread adoption of beacons in smart cities also means they are common targets of adversaries with economic or political motives.

Compared to typical wireless communication, a beacon is a resource-constrained device transmitting a fixed, static message to multiple receiving devices via a one-way, broadcast channel, posing several challenges to securing beacon systems. Broadcast devices are susceptible to tracking and message spoofing as the messages can be received by all nearby devices (including attackers' devices). Since the same message is always broadcasted by a beacon, this eases attackers' job to clone the message and makes it harder for legitimate users to differentiate between authentic and fake beacon messages. For example, digitally signing a beacon message cannot withhold a replay attack wherein an attacker eavesdrops the beacon's signed message and replays it easily. In addition, the resource constraints of beacons also limit the range of possible defence mechanisms that can be deployed. Finally, beacons are typically used in context awareness applications, often with a cyber-physical dimension. The message itself carries no useful information without interpretation against the physical context. Attackers often aim to mess up the contextual data while delivering a cloned beacon message to users. Defence considerations should therefore cover beyond the common goal of wireless communication (i.e. to assure message privacy and integrity).

---
[1] Aldar C-F. Chan is with University of Hong Kong, aldar@graduate.hku.hk.
[2] Raymond M. H. Chung is with University of Hong Kong, manhon@graduate.hku.hk.

This article gives an exposition of the attack surface of BLE beacon systems, enumerating vulnerabilities and characterizing attacks based on attacker goals, attacker capabilities required to launch an attack and the impact of a successful attack. Such a framework would provide practical guidelines for application developers to identify security issues of their beacon deployments and understand potential attacks against them so that proper protection can be formulated based on the desired security goals. Since the discussed attacks are based on high-level operations of BLE communication, the attacks are of the same nature, though to a different degree, in newer BLE protocols; the attack characterization is therefore equally applicable.

The rest of the article is organized as follows. The next section presents the operation of beacon systems. Section 3 briefly introduces different attacks against beacon systems, with a characterization of attacks given in Section 4. Section 5 introduces possible protection mechanisms and Section 6 illustrates how the attack characterization table can be used practically. Section 7 discusses future trends.

## 2. Operation of BLE Beacon Systems

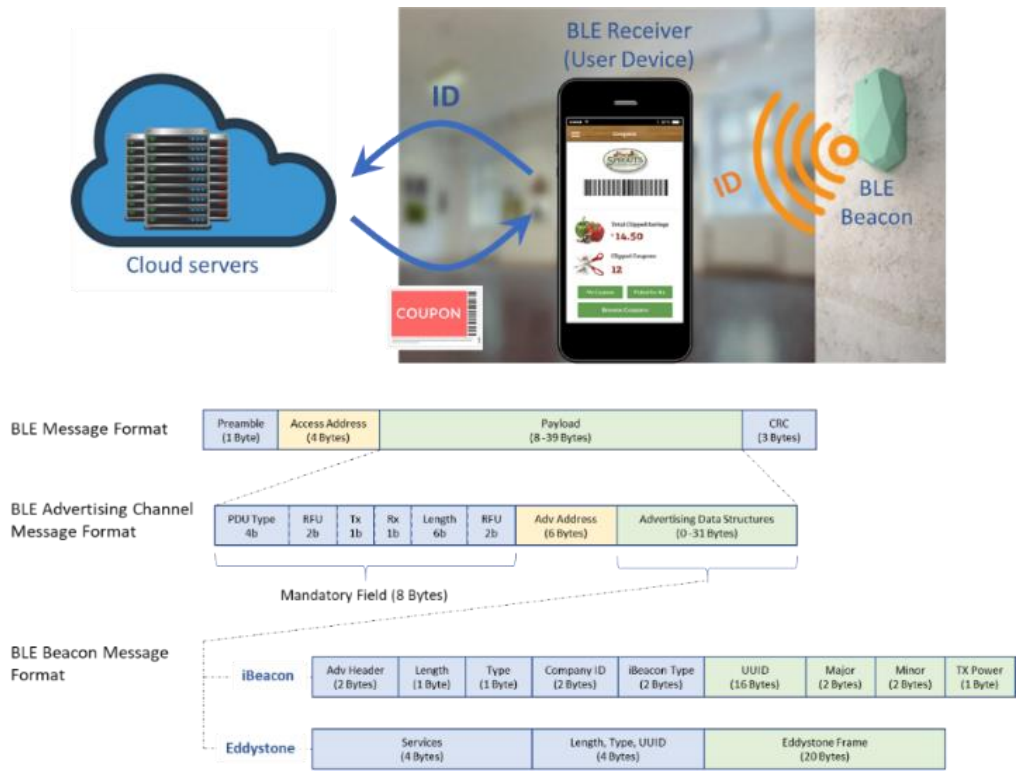

Figure 1. Typical operation of a BLE beacon system and beacon message formats.

A beacon differs from usual BLE devices in that it only has a one-way channel for broadcasting its predefined identity (ID) for other mobile devices to pick up. Figure 1 shows the operation of a typical beacon system. Beacons usually work with a dedicated smartphone application. A beacon is programmed to broadcast a fixed ID at regular intervals, thus making beacon cloning easy, simply through passive eavesdropping and ID copying. When a smartphone pre-installed with the application is in proximity of a beacon and successfully receives this ID, the application would look up from this ID relevant actions that should be taken if it can recognize the ID. The action usually includes downloading a piece of content stored on a cloud or backend server and displaying it on the smartphone. The beacon message itself carries no content and has to rely on the installed application to look up from the ID the actual location or URL for downloading relevant content. In this way, a shop with beacons installed could automatically push marketing content to a nearby smartphone. If an adversary manages to interfere the mapping between beacon IDs and the locations of the referred content, wrong content will be delivered to cause confusion. Attacks presented in the next section largely belong to this type.

As depicted in Figure 1, all beacon messages are carried in the same advertising data structure of BLE regardless of the manufacturers [3]. Yet, there are different message formats for beacons from different manufacturers. The choice of the message format may involve different trade-offs. In contrast to connected BLE sessions with multiple security mechanisms in place, BLE advertising transmissions used by beacons are unprotected, thus making cloning easy.

## 3. Security Issues and Privacy Concerns of BLE Beacon Systems

There are various security issues regarding beacon systems, including piggybacking, spoofing, silencing, user device resource draining, re-programming and reshuffling. Besides, beacons are commonly used for location-based or proximity-based services [6, 9, 14]. These services are exposed to possible privacy breaches [5, 6, 8-10, 12-14], including user profiling [9, 14] for the former and presence inference [6, 9] for the latter.

*Piggybacking*

A piggybacking attack [9, 14] (also called a hijacking attack in [9]) refers to the attack wherein an adversary wants to run its service leveraging on the beacon infrastructure built by others without their consent. Since each beacon ID is broadcast in clear, an attacker can listen to all beacon IDs of interest and build a database of beacon IDs and their locations and other details, with which he can build his services using these third-party beacons. In the worst scenario, the attacker could be a business competitor of the beacon owner, so that he can offer a coupon with a more attractive deal for customers whenever the customers receive a coupon from the beacon owner as triggered by the proximity to the beacons.

*Spoofing*

After listening and recording a beacon ID and other details, it is possible for an attacker to forge a fake beacon with the same ID (possibly placed in different geographic locations) by spoofing the original beacon ID [14]. Users cannot tell apart a real beacon and a fake beacon since their IDs are the same. It should be noted that while message authentication (such as digital signatures) typically works for assuring integrity of a wireless communication channel, it cannot withhold spoofing attacks against a beacon system. An attacker could record and replay beacon messages and their signatures, which are indistinguishable from those of a real beacon. A fake beacon, however, could cause confusion, especially when used for providing location-based services.

*Silencing*

A silencing attack [9] aims to manipulate a smartphone application and deceive it into perceiving a beacon as remote by using cloned beacons or BLE transmitters to transmit a flood of spoofed beacon messages with a greater TX Power value. As such, the application's estimation of proximity would become biased towards the fake readings and result in an inflated estimated value for the distance between the real beacon and application. Consequently, the application would not show the content or notification as it should for being in proximity of the beacon, therefore undermining service availability.

*User Device Resource Draining*

The goal of a resource draining attack, which is also called user harassment in [9], is to slow down user devices or make other services on them unavailable. By broadcasting many different beacon IDs, a nearby smartphone with an installed beacon-aware application would keep comparing and processing these IDs to an extent that computing resources available for other applications are considerably reduced. The smartphone's battery may also be exhausted.

*Re-programming*

Some beacons allow over-the-air programming or software update, without proper authentication implemented due to cost and power consumption constraints. An attacker can possibly program these beacons with new IDs [10]. This would make the original content service associated with the beacons unavailable, which is a kind of denial of service attacks. In addition, by carefully crafting the new IDs, an attacker can cause wrong content to be delivered to user applications as in spoofing attacks. Hence, re-programming attacks can possibly undermine both integrity and availability of a beacon system.

*Reshuffling/Cracking*

Since proximity awareness is a key characteristic of beacon-aware applications, exposure of beacons to users seems inevitable in most deployment scenarios. Beacons are usually placed to cover an area and physically accessible to all users including potential attackers. If an attacker manages to swap the positions of beacons physically, the mapping between beacon IDs and the URLs or addresses of content (which is programmed on a user application) would be wrong [4], thus leading to wrong content to be downloaded for a misplaced beacon. If physical removal of beacon by an attacker occurs, the existing beacon service would become unavailable to user applications [14].

*User Profiling*

A user profiling attack usually tracks locations of a user through an application installed on the user's device [9, 14]. The adversary could be the developer of an authorized beacon-aware application which the user has willingly installed, but the developer has not obtained consent from the user to track his location history. Since the two set of API services — needed by the authorized application to provide beacon-enabled services and used for illegitimate user profiling — usually have a large overlap, it is generally difficult for users to detect the unlawful surveillance functionality hidden in the authorized application. In other cases, the adversary could even deceive a user into granting service authorization to a tracking application and free-ride on third-party beacons to track the user's locations [9]. As the installed application listens to beacons in proximity, it will upload these beacon IDs to a dedicated server hosted by the adversary for storage and lookup of locations of the beacons. User profiling is not directly attributed to exploitation on beacon systems but due to installation of a malicious application leveraging the beacon infrastructure.

*Presence Inference*

A presence inference [5-6] attack targets at beacon-emitting objects carried by users, with the adversary installing surveillance equipment in certain areas to detect messages emitted from the target users' beacons to infer their presence in the areas.

# 4. Characterization of Attacks against BLE Beacon Systems

Table 1 shows a detailed characterization of potential threats against typical BLE beacon systems, in terms of the adversary motives, requirements on adversarial capabilities needed to launch a successful attack, resulting impact of a successful attack on beacon infrastructure owners and users, and applicability of different defence mechanisms to these attacks. This characterization framework could be useful for understanding and evaluating security issues of a given deployment in two aspects. First, it helps systematically assess the difficulty to launch a potential attack against the deployment based on an attacker's investment needed to launch the attack and how likely an attacker has the necessary skills, information and device/equipment setup required to launch the attack. Second, it helps estimate the potential impact of a successful attack on the beacon infrastructure owner and users. In addition, it gives practical guidelines on the applicability of different defence strategies to the concerned attack. As an example, if it is relatively easy to launch a certain attack and a successful attack could possibly cause great impact on the owner or users, weight should be given to the consideration of investing in one of the defence mechanisms to safeguard the beacon system. The details of how to use Table 1 is given in Section 6. Based on the characterization, attacks against beacon systems are classified in Figure 2.

There are several attacks against a BLE beacon system, some of which target at the infrastructure owner (including piggybacking, silencing, spoofing, reprogramming, reshuffling/cracking attacks) and other at the users (including user profiling, presence inference and user device resource draining attacks). User profiling and presence inference mainly aim to track valid users, or deduce their habits using a beacon infrastructure, which is concerned with user privacy. The other attacks largely aim at a beacon infrastructure per se to allow an attacker to free-ride on its service or render its service unusable or unavailable to cause annoyance to legitimate users.

## 4.1) Attacker Motives

While there are different types of threats associated with a beacon system, the adversary objectives seem rather focused, with roughly three types of motives:

1) *Free riding (M1)* — an attacker aims to use an existing beacon infrastructure established by others for free (piggybacking attacks) to push his selected content or notifications to its users;

2) *User profiling (M4)* — an attacker aims to use an existing beacon infrastructure as a tool for tracking its users or deduce their habits (user profiling and presence inference attacks); and

3) *Service disruption (M2, M3, M5)* — an attacker aims to disrupt the service of a beacon system to cause annoyance to users of beacon-aware applications (including silencing, spoofing, reshuffling/cracking, re-programming and user device resource draining attacks) that may lead to the removal of the applications, thereby undermining reputation of the respective corporates.

A free-riding attacker could be a company which wants to develop an application to push location-based notifications or content to potential customers but does not want to invest in establishing the needed beacon infrastructure.

An attacker aiming to profile users tracks persons of interest and infers their habits with beacons. However, large-scale surveillance would still be challenging [6, 9]. Practically, it is possible to construct — using a user profiling attack — a very accurate profile of a targeted user, including details such as how often he visits a retail chain, which departments he prefers and how much time he spends in front of a particular product, etc. Such data would be valuable for personalized advertising [9].

Table 1. Characterization of different types of attacks against a BLE beacon system

| Type of attack | Adversary's motives | Breached security goal (CIAP) | Required adversarial capabilities Level of skillset required: High (H), Medium (M), Low (L) | | | | | | | Impact to legitimate users Impact level: High (H), Medium (M), Low (L) Affected party: User (U), Owner (O) | Possible defence strategies | | |
|---|---|---|---|---|---|---|---|---|---|---|---|---|---|
| | | | C1 | C2 | C3 | C4 | C5 | C6 | C7 | | Time-varying ID | Outlier detection | Access jamming |
| | | | L | L | M | H | H | H | M | | | | |
| Piggybacking /Hijacking Attack (A1) | • Free riding of beacon infrastructure built by others for self-use (M1) | C | √ | √ | | | | | √ | • Limited impact to users (L, U) <br> • Loss of revenue to beacon infrastructure owners (H, O) | • | | |
| Spoofing Attack (A2) | • Messing up the ID-content mapping of an existing beacon infrastructure (M2) | I | √ | | √ | | | √ | | • Wrong information content delivered to users (H, U) <br> • Reputational risks to beacon infrastructure owners (M, O) <br> • Corrective action by infrastructure owners required (H, O) | • | • | |
| Silencing Attack (A3) | • Disabling part of an existing beacon infrastructure (M3) | A | √ | | √ | | | √ | | • Beaconing service in affected areas unavailable to users (H, U) <br> • Reputational risks to beacon infrastructure owners (M, O) | • | | |
| Re-programming Attack (A4) | • Messing up the ID-content mapping of an existing beacon infrastructure (M2) <br> • Disabling part of an existing beacon infrastructure (M3) | I/A | | | | √ | | | | • Wrong information content delivered to users (H, U) <br> • Beaconing service in affected areas unavailable to users (H, U) <br> • Reputational risk to beacon infrastructure owners (M, O) <br> • Corrective action by infrastructure owners required (H, O) | | • | |
| Reshuffling /Cracking Attack (A5) | • Messing up the ID-content mapping of an existing beacon infrastructure (M2) <br> • Disabling part of an existing beacon infrastructure (M3) | I/A | | | | | √ | | | | | • | |
| User Profiling (A6) | • Tracking of user locations/activities (M4) | P | √ | √ | | | | | √ | • Privacy of users breached (H, O) <br> • Locations and activities of users leaked to unauthorized parties (H, U) | • | | |
| Presence Inference (A7) | • Tracking of user locations/activities (M4) | P | √ | √ | | | | √ | | | • | | • |
| User Device Resource Draining (A8) | • Exhausting/disabling user devices (M5) | A | | | | | | √ | | • Processing capability of user devices reduced (M, U) <br> • Services on user devices may become unavailable to users (H, U) | | | |

*Required adversary capabilities*
C1: Passive eavesdropping and recording of beacon IDs
C2: Reconstructing beacon ID database(s)
C3: Cloning or production of fake beacons
C4: Read/write access to beacon firmware (remotely or physically)
C5: Physical access to beacons to remove or reshuffle them to new locations
C6: Installation of beacons (fake or new) and/or surveillance equipment in areas of interest
C7: Ability to get users to install the adversary's application and grant required authorizations

An attacker with a service disruption motive aims to bring down service of a beacon infrastructure to make it unavailable to users or disrupt the established mapping between beacon IDs and content addressing information to cause wrong content to be delivered to users' applications. For example, signages or navigation instructions can be wrongly delivered or simply unavailable to a user application whenever needed. Consequently, users may find it annoying and therefore remove the application. This could hurt credibility of the respective corporation.

While a free-riding or user profiling attack would be less disruptive to users, an attack aiming to mess up the beacon ID mapping or make beacon service unavailable could possibly bring great confusion and impact on users.

By viewing the beacon broadcast as a communication channel, attacks and attacker motives can be mapped to the conventional CIA (Confidentiality, Integrity, Availability) security goals of typical information systems. In order to free-ride on a beacon infrastructure, an attacker has to be able to read and interpret its beacon messages. This is equivalent to breaching message confidentiality of the beacon channel. An attacker able to mess up the mapping between beacon IDs and content addressing information violates message integrity of the beacon channel. An attacker able to bring down the service of a beacon infrastructure undermines the availability goal of the beacon channel. Yet, privacy (P) should be considered as a separate goal from confidentiality as the message itself may not have to be revealed for breaching privacy.

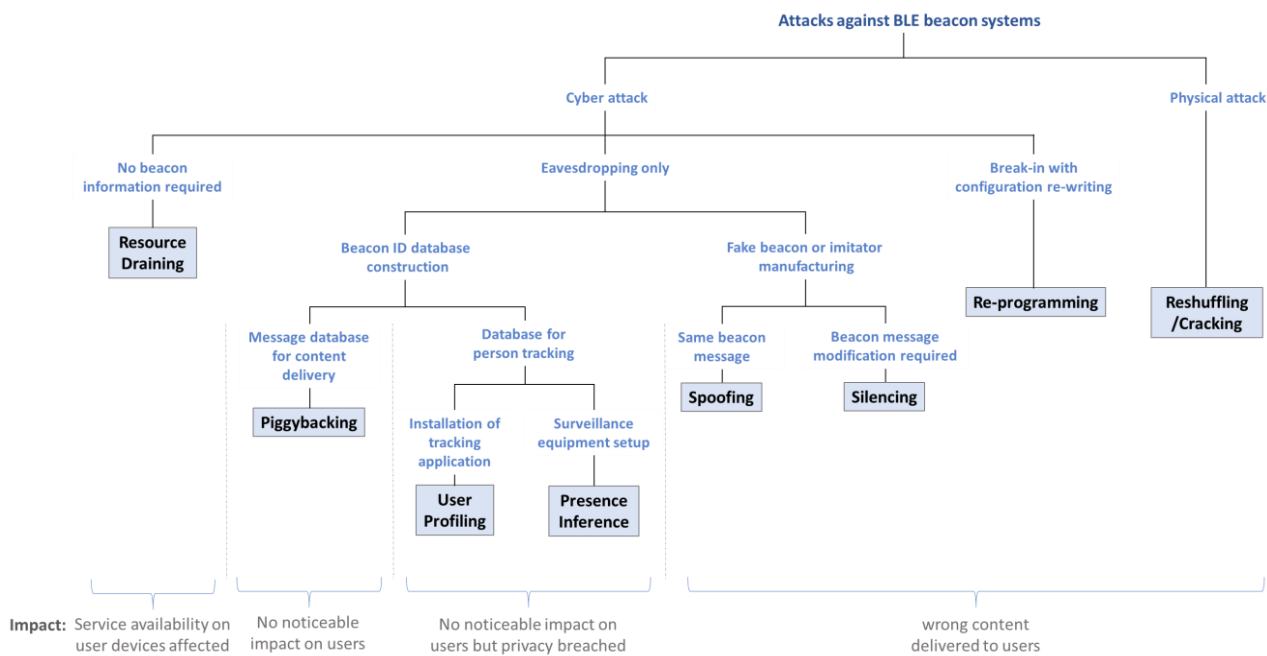

Figure 2. Classification of attacks against a beacon system

*4.2) Threat Model / Adversary Capability*

Seven adversarial capabilities which allow an attacker to launch various attacks against a beacon system are identified in Table 1. The level of skillset required of the adversary corresponding to these capabilities is rated as low (L), medium (M) and high (H). These attacks usually require an attacker to either intrude beacon devices (re-programming and reshuffling attacks) or, less intrusively, build an overlay coexisting with a beacon infrastructure (silencing, spoofing, presence inference and resource draining attacks). Others (piggybacking and user profiling attacks) only require knowledge of IDs and locations of beacons to leverage on an existing beacon infrastructure. For the case of intrusion, it could be remote (as in re-programming) that an attacker has read/write access to beacon firmware due to poorly secured beacons (Capability-C4) or physical (as in reshuffling/cracking) that an attacker can physically remove or reshuffle beacons to new locations (Capability-C5). Beacons are usually hidden in different spots to cover an area, thus rendering them easily accessible for an attacker to remove them, open them up or relocating them. The intrusive attacks are realistic threats for typical deployments.

For the less intrusive attacks, an attacker is always assumed to possess software and hardware to allow him to sniff BLE packets including beacon messages to learn IDs of different beacons (Capability-C1). Such hardware is typically inexpensive, and the software usually free. This presumes that the attacker can come to proximity with the victim. Yet, an attacker may also use high-gain antennas to allow him to eavesdrop or imitate beacons from a larger-than-typical distance [9]. This allows an attacker to capture and replay packets or even craft and inject new ones with modified data fields. It is fair to assume that an attacker has knowledge of IDs and locations of beacons of interest. This knowledge is necessary for an attacker to rebuild a beacon ID database (Capability-C2) in order to launch the piggybacking and user profiling attacks, or clone beacons (Capability-C3) for silencing and spoofing attacks. Besides cloning beacons, an attacker may also use a BLE transceiver to replay messages of different beacons to imitate them to deceive users.

To disrupt beacon service in order to annoy legitimate users, an attacker needs to install his cloned beacons or equipment to imitate beacons in areas of interests (Capability-C6) in order to create the wrong mapping to cause wrong content to be delivered. In a silencing attack, a cloned beacon also needs to be close to the real beacon to deceive smartphone applications. Similarly, to achieve presence inference, an attacker needs to install his own surveillance equipment in an area where the person of interest would have a high chance to appear in order to detect his beacon tags.

In some cases, an attacker must be able to develop a beacon-aware application and distribute it as legitimate such that users would grant authorization to the required API services (such as CoreLocation services on iOS) to this application (Capability-C7). In piggybacking attacks, the application is indistinguishable from an authentic application since it appears as a common beacon-enabled application. The only difference is that the application uses a third-party beacon infrastructure built by another party without his consent. In contrast, user profiling attacks need to deceive users into installing a malicious application which secretly monitors a user's vicinity of third-party beacons.

### *4.3) Impact of Attacks*

The actual impact of an attack is closely related to the adversary motives. While attacks aiming at free-riding or user profiling would have no visible impact to users — though user privacy is breached sneakily in user profiling, those targeting at service disruption would cause annoyance to users and normally require prompt remediation by the beacon infrastructure owner. The level of these impacts to the affected parties (i.e. User (U) and Owner (O)) is classified as High (H), Medium (M) and Low (L).

## 5. Possible Defence Strategies

Most attacks require knowledge of beacon IDs and their locations. Correspondingly, there are two approaches for an attacker to collect beacon IDs, namely, the lunch-time attack and pervasive eavesdropping. In a lunch-time attack, an attacker collects all beacon IDs initially in a collection phase and then it would be too costly for him to do so afterwards if the beacon owner makes changes during deployment. In pervasive eavesdropping, an attacker is able to listen to changing beacon IDs on the fly without constraints. It is most difficult to defend against attackers with this eavesdropping capability, and most of the existing defence mechanisms only assume a lunch-time adversary.

It is challenging to implement defence for beacons because they are usually equipped with a unidirectional channel which makes conventional authentication protocols (which require interaction) inapplicable. Simply appending a digital signature to a beacon ID would not work to withhold spoofing attacks. It has to combine with a changing ID. *In general, there are three strategies to defend beacons against attacks, with the mapping given in Table 1.*

***Time-varying ID [7, 10].*** Beacons are programmed to broadcast evolving or time-varying IDs, thus making it more costly for an attacker to copy and replay beacon messages. Industrial efforts, such as the Gimbal and Kontakt beacons, have also come up with a similar but less secure version based on rotating IDs from a predefined, fixed pattern. In more secure versions [7, 10], cryptographic techniques such as keyed pseudorandom functions are used to generate time-varying beacon IDs. To overcome synchronization issues between beacons and user applications, [10] applies time as input to the pseudorandom function for generating beacon IDs and uses a Bloom filter to accommodate loose time synchronization. However, generation of evolving IDs could reduce the lifespan of the battery of a beacon. In addition, evolving IDs cannot withhold re-programming and reshuffling attacks.

*Outlier/Anomaly Detection [4, 11].* Traces of user queries are gathered at the backend server in order to run hypothesis testing to detect any outlier or anomaly of beacon ID transitions with respect to their geographic mounting positions. As a user device moves around a beacon-equipped area, the transitions of consecutive beacon IDs seen by the application would follow certain probability distribution (Figure 3). [4, 11] models these transitions with a Markov chain and derive the transition probabilities based on the relative mounting positions of beacons. A merit of outlier detection is that it requires no modification on beacons and does not shorten beacon lifespan. While outlier detection is more powerful and can defend all attacks other than piggybacking attacks, careful consideration of user privacy protection is necessary.

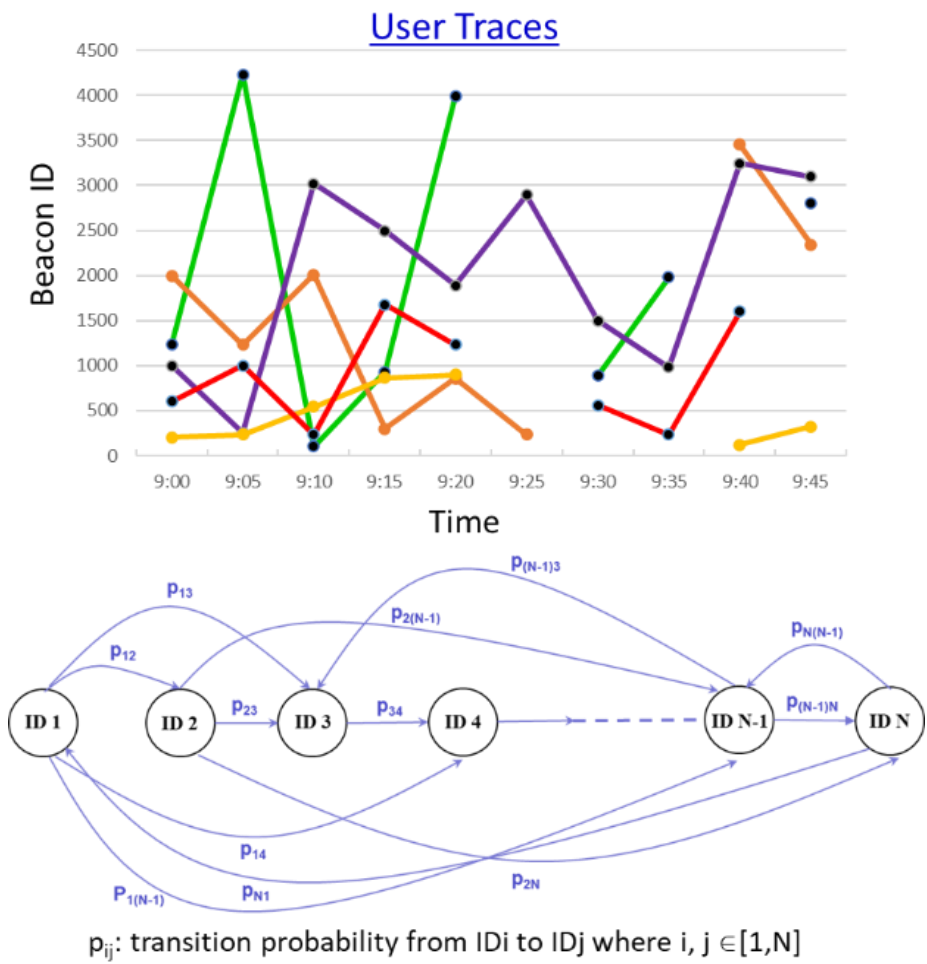

Figure 3. Traces of user queries with beacon IDs modelled as a Markov chain.

*Selective Jamming [6].* An additional BLE-enabled guardian device can be installed to opportunistically invoke reactive jamming of messages emitted by a personal beacon tag. Only devices authenticated through an out-of-band channel with the guardian device can read beacon messages indirectly through the guardian device. No modification on beacons is required. But this approach mainly applies to protecting beacon tags carried by a person against presence inference and would not be useful for beacons deployed for location-based services.

## 6. Security Evaluation and Defence Formulation

Figure 4 shows how Table 1 (which characterize attacks) can be used to assess the likelihood of an attack and formulate the respective defence strategy. As the first step (Step 1a), all the adversarial motives (in the second column) that are probable based on the application scenario and local context are selected. This corresponds to a number of attacks which an adversary has sufficient incentive to launch. For example, under a certain social situation with observable protests against contact tracing, there is high incentive to disrupt the respective beacon service and make it unavailable (motivations M2, M3); the adversary is incentivized to launch attacks A2, A3, A4 and A5.

Then, in Step 1b, all the applicable capabilities of the potential adversary are selected from respective "adversarial capability" columns (based on an understanding of the adversary's competence), corresponding to several attacks which the adversary is capable to launch. For example, the adversary has capabilities C1, C2, C3 and C6 and is therefore capable to launch attacks A2, A3, A6 and A7.

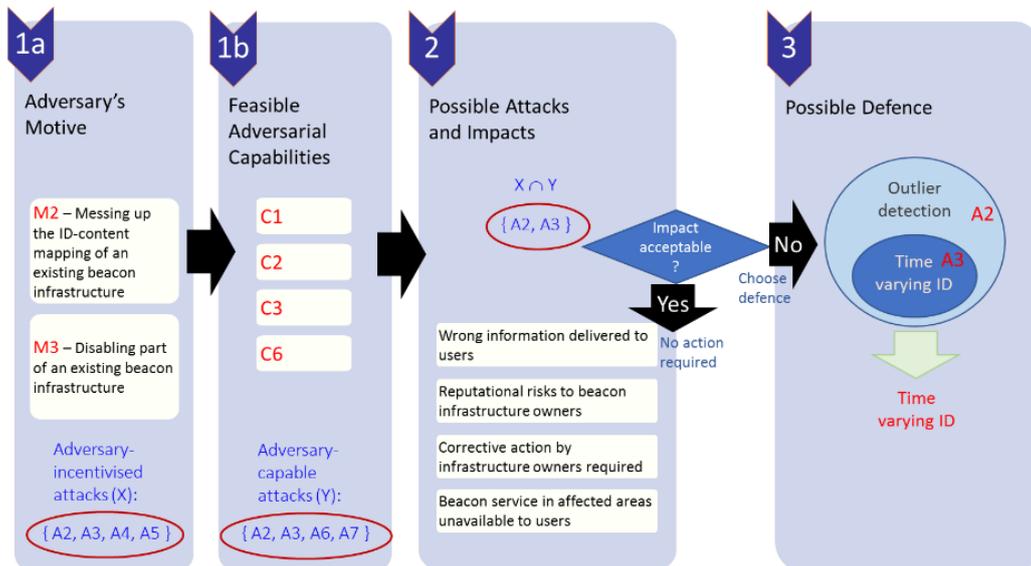

Figure 4. Steps to use the attack characterization table.

*The likelihood of an attack depends on whether an adversary has the incentive and whether he is capable to launch it.* Hence, the intersection of the two sets of attacks obtained in Step 1a and 1b is the set of attacks with a high likelihood to occur. In Step 2, the impacts of these attacks can be read out from Table 1. Depending on whether the impact is tolerable, defence strategy formulation is required or can be skipped.

In Step 3, for each attack identified in Step 2, all the applicable defence mechanisms are read out from Table 1. The common intersection of these sets of mechanisms (corresponding to all the identified attacks) is the set of applicable defence mechanisms to safeguard the beacon infrastructure against all the identified attacks. For example, attack A2 can be defeated by time-varying ID or outlier detection, whereas, only time-varying ID is applicable to defend against attack A3; hence, only time-varying ID is applicable to safeguarding against both attacks.

## 7. Future Trends

As new generations of beacon devices (such as Apple's AirTag) become more computationally powerful, more sophisticated cryptographic defence could be implemented in the beacons, say, for ID rotation. However, defence mechanisms on the receiver side are flexible and therefore expected to be mainstream. Besides, additional data with increasing variety are increasingly deployed in emerging location-based and proximity-based services. Harnessing these new contextual data, more accurate outlier detection can be achieved. Newer algorithms to combine these data streams are expected to emerge.

## 8. Conclusions

While the benefits of BLE beacon systems for smart city applications are clear, they can be undermined by the inherent security and privacy risks of beacon systems. These risks primarily stem from the unprotected nature of fixed beacon messages over a broadcast channel. This article enumerates various security and privacy attacks against beacon systems to characterize them based on adversary motives and capabilities, impact of a successful attack, and effectiveness of known defence techniques. While various attacks exist, the adversary has rather focused objectives, including free-riding, service disruption and person profiling. The impact of service disruption on beacon system owners and users is generally high, whereas there is no noticeable impact of free-riding and person profiling despite that the latter sneakily breaches user privacy. There are a handful of techniques to protect beacon systems involving various trade-offs.

**Reference**

[1] Apple. iBeacon specifications, accessible at https://developer.apple.com/ibeacon.

[2] L. Atzori, A. Iera, and G. Morabito, "From 'Smart Objects' to 'Social Objects': the next evolutionary step of the Internet of Things," IEEE Communication Magazine, 52(1): 97-105, Jan, 2014.

[3] Bluetooth Special Interest Group. Specification of the Bluetooth system, ver. 4.2, accessible at: https://www.bluetooth.com/specifications.

[4] A. C-F. Chan, and M. H. Chung, "System and method for attack detection in wireless beacon systems," US Patent 10,699,545 B1, granted, 30 Jun, 2020.

[5] A. K. Das, P. H. Pathak, C.-N. Chuah, and P. Mohapatra, "Uncovering privacy leakage in BLE network traffic of wearable fitness trackers," in proc. 17th Int. Workshop on Mobile Computing Systems and Applications, p.99-104, 2016.

[6] K. Fawaz, K.-H. Kim, and K. G. Shin, "Protecting privacy of BLE device users," in proc. USENIX Security 16, p.1205-1221, 2016.

[7] A. Hassidim, Y. Matias, M. Yung, and A. Ziv, "Ephemeral identifiers: Mitigating tracking & spoofing threats to BLE beacons," 2016.

[8] K. E. Jeon, J. She, P. Soonsawad, and P. C. Ng, "BLE beacons for Internet-of-Things applications: survey, challenges and opportunities," IEEE IOTJ, 5(2): 811-828, Apr, 2018.

[9] C. Kolias, L. Copi, F. Zhang, and A. Stavrou, "Breaking BLE beacons for fun but mostly profit," in proc. EuroSec'17, 2017.

[10] A. Lam, F. Wong, and A. Chan, "Secure BLE broadcast system for location based service," US Patent 10,219,106 B1, granted, 26 Feb, 2019.

[11] C. Liu, P. Zhao, K. Bian, T. Zhao, and Y. Wei, "The detection of physical attacks against iBeacon transmitters," in IEEE/ACM 24th Int. Symp. Quality of Service (IWQoS), 2016, pp. 1–10.

[12] J. Priest, and D. Johnson, "Covert channel over Apple iBeacon," in proc. SAM, p.51, 2015.

[13] M. Ryan, "Bluetooth: With low energy comes low security," in proc. WOOT'13, p. 4, 2013.

[14] P. Spachos, and K. Plataniotis, "BLE beacons in the smart city: applications, challenges, and research opportunities," IEEE IOTM, 3(1): 14-18, Mar, 20